\documentclass[twocolumn,showpacs,preprintnumbers,amsmath,amssymb]{revtex4}
\usepackage{graphicx}
\usepackage{dcolumn}
\usepackage{bm}
\usepackage{comment}
\usepackage{subfigure}
\usepackage{amsmath}
    \DeclareMathOperator{\sech}{sech}
\usepackage{url}
\usepackage{hyperref}
\usepackage{breakurl}

\begin{document}

\title{The effects of ion mass variation and domain size on octupolar out-of-plane magnetic field generation in collisionless magnetic reconnection} 
\author{J. Graf von der Pahlen and D. Tsiklauri}
\affiliation{School of Physics and Astronomy, Queen Mary University of London, London, E1 4NS, United Kingdom}
\date{\today}
\begin{abstract}
 J. Graf von der Pahlen and D. Tsiklauri, Phys. Plas. 21, 060705 (2014), established that the generation of octupolar out-of-plane magnetic field structure in a stressed X-point collapse is due to ion currents. The field has a central region, comprising of the well-known qaudrupolar field (quadrupolar components), as well as four additional poles of reversed polarity closer to the corners of the domain (octupolar components). In this extended work, the dependence of the octupolar structure on domain size and ion mass variation is investigated. Simulations show that the strength and spatial structure of the generated octupolar magnetic field is independent of ion to electron mass ratio. Thus showing that ion currents play a significant role in out-of-plane magnetic structure generation in physically realistic scenarios. Simulations of different system sizes show that the width of the octupolar structure remains the same and has a spacial extent of the order of the ion inertial length. The width of the structure thus appears to be independent on boundary condition effects. The length of the octupolar structure however increases for greater domain sizes, prescribed by the external system size. This was found to be a consequence of the structure of the in-plane magnetic field in the outflow region halting the particle flow and thus terminating the in-plane currents that generate the out-of-plane field. The generation of octupolar magnetic field structure is also established in a tearing-mode reconnection scenario. The differences in the generation of the octupolar field and resulting qualitative differences between $X$-point collapse and tearing-mode are discussed.
\end{abstract}


\maketitle

\pagestyle{plain}
\thispagestyle{plain}

\section{Introduction}
\pagestyle{plain}
\thispagestyle{plain}

Magnetic Hall reconnection, first proposed by B. Sonnerup \cite{sonnerup}, is a mode of reconnection relying on the decoupling of ions and electrons in a diffusion region and is of great interest in the study of magnetic reconnection. It presents an alternative to the the Petschek model, which relies on an anomalous resistivity \cite{Erkaev}. Even in setups suitable for Petschek reconnection, contributions of Hall effects need to be considered. A recent analytical result, corroborated by a numerical study, shows that the transition from Petschek to Hall reconnection occurs when the half-length of the current sheet reaches the ion inertial length \cite{arber}. This was shown to be a direct consequence of a generalised scaling law, relating the reconnection rate to the distance between the $X$-point and the start of slow mode shocks.

An observational consequence of Hall reconnection is the generated quadrupolar out-of-plane magnetic field, induced by currents resulting from the decoupling of electrons from ions i.e. the Hall currents, first demonstrated in a study by Teresawa \cite{terasawa}. The effect was further shown to occur in numerical Hybrid simulations \cite{mandt,karimabadquad,arznerquad} and later in a full Particle In Cell (PIC) numerical simulation \cite{prittchet}. However, as shown in Ref. \cite{karimabadioct}, kinetic simulations of magnetic reconnection, where the Hall term was excluded, can also lead to quadrupolar magnetic field structure generation, due to ion diamagnetic drifts driven by an anisotropic ion stress tensor. By being an observational signature of magnetic reconnection, the quadrupolar field has thus been of great interest in recent  spacecraft missions, including Polar \cite{Mozer} and Cluster \cite{Borg}. Both missions observed individual magnetic poles in the magneto-tail of the Earth. Subsequently, a full quadrupolar pattern was observed in a multi-spacecraft Cluster mission \cite{Eastwood}. The experimental evidence of the full quadrupolar structure was also found at the MRX facility \cite{mrx}. However, it was shown in Ref. \cite{karimabadi.other.quadrupoles} that inhomogeneous ion flow and pre-existing out-of-plane magnetic fields, can lead to the generation of quadrupolar out-of-plane magnetic field structure without the Hall term. Thus, the generation of a quadrupolar magnetic field is not necessarily a tell-tale sign of Hall-mediated magnetic reconnection.

Ref. \cite{Uzdensky} proposes an analytical model, explaining the Hall out-of-plane quadrupolar magnetic field near an $X$-point as the result of electron motion towards and away from the $X$-point, as field lines reconnect. A uniform ion distribution was assumed. Since ions decouple from the magnetic field sooner than electrons, they move independently of field lines near the $X$-point. On the other hand, electrons are assumed to be coupled to the field lines and thus only move with the field and along the field lines. Since the spacing of field lines increases as they approach the $X$-point, the electron density decreases. Thus, due to the uniform ion density, this results in a net positive charge. Electrons in the inflow region therefore move along the field lines to towards the $X$-point to restore charge neutrality and then move away from the $X$-point in the out flow region, leading to a quadrupolar pattern.

  \begin{figure}[htbp]
    \includegraphics[width=1.0\linewidth]{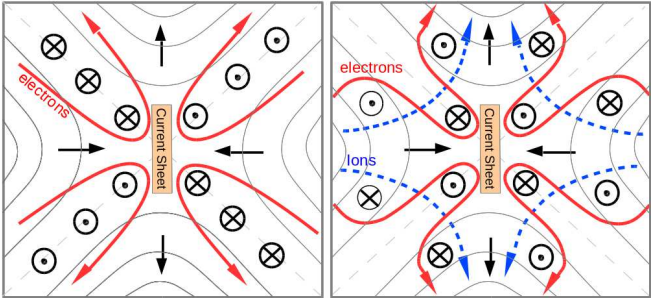}
     \caption{ \label{Fig:model}(Left panel) Reconnection at an $X$-point superimposed with electron motion (as indicated by the labelled track) and the resulting out-of-plane magnetic field structure, as given by the analytical model in Ref. \cite{Uzdensky}. Black arrows on field lines signify inflow and outflow regions. Ions here are assumed to be decoupled and uniformly distributed. As the spacing of field lines increases at the $X$-point, coupled electrons in the inflow region move towards the $X$-point to restore charge balance. Similarly, the electrons in the outflow region move outwards along the field lines and the characteristic quadrupolar out-of-plane magnetic field structure is generated. Due to the non-inclusion of ions, out-of-plane magnetic structure is not localised and extends along the seperatrices. (Right panel) The setup and resulting current and out-of-plane magnetic field generation for $X$-point collapse, from simulations in Ref. \cite{gvdp}. It is shown that, as in-plane field lines reconnect, electrons move towards and then away from the $X$-point, generating quadrupolar structure as described in Ref. \cite{Uzdensky}. Away from the $X$-point, ions move independently from the field (see dashed tracks), generating a magnetic field of opposite polarity to that of the quadrupole, thus generating an overall octupolar structure. Electrons at the edge of the ion diffusion region move such that they cancel the out-of-plane field, thus making it \textit{localised}.} 
   \end{figure}

While their model describes the generation of the quadrupolar field, Ref. \cite{Uzdensky} points out that this model is limited by the non-inclusion of ion currents, resulting in a quadrupolar field that stretches along separatrix arms indefinitely. In a recent simulation study by J. Graf von der Pahlen and D. Tsiklauri \cite{gvdp} it is shown that, in an $X$-point collapse scenario, ion currents not only provide a cut-off to the quadrupolar field, but contribute to the out-of-plane magnetic structure themselves. The out-of-plane magnetic field that emerged in Ref. \cite{gvdp} was shown to have the well-known quadrupolar field at the centre, generated by electron currents, and four regions of opposite magnetic polarity on the outside, resulting from ion currents, as illustrated in Fig.~\ref{Fig:model}. The overall field appears to have an octupolar structure (we shall refer to the inner magnetic quadrupolar field as quadrupolar components and the outer field of opposite polarity as the octupolar components). 

At the beginning of the simulation, a different type of octupolar magnetic field, with smaller field strength ($\approx 3\%$ of external in-plane field),  emerges (see Fig. 2 in Ref. \cite{gvdp}). The same effect was demonstrated in Ref. \cite{karimabadioct} (see their Fig. 5) using a hybrid simulation. However, the present study focusses on the larger field structure ($\approx 15\%$ of external in-plane field), emerging \textit{later} in the simulation. Further, a octupolar signatures have been observed in tearing-mode reconnection scenarios with multiple islands as shown in Ref. \cite{karimabadi.other.octupoles}, Fig. 4, and in Ref. \cite{pritchett.other.octupoles}), Fig. 3. The emergence octupolar structure in these scenarios is linked to the island coalescence and is beyond the scope of this paper.

Here, we extend Ref. \cite{gvdp} by investigating the dependence of the octupolar field on variation of electron to ion mass ratio and the domain size. It is established that the generation of octupolar field is neither the result of unphysical boundary conditions nor unrealistic mass ratios. Thus we show that the results are relevant for real laboratory experiments and spacecraft observations. Further, by simulating a tearing-mode set-up, as previously studied in Ref. \cite{prittchet}, it is shown that a similar type of octupolar structure can also be found in reconnection scenarios other than $X$-point collapse, where their emergence have been previously overlooked.

\section{Simulation Setup}
\pagestyle{plain}
\thispagestyle{plain}

Previous works on collisionless $X$-point collapse can be found in Refs. \cite{tsiklauri2008,Tsiklauri2007,gvdp1,gvdp}. In the simulation results presented in this study, the in-plane magnetic field is that of a standard $X$-point collapse configuration, first introduced by Dungey in 1953 \cite{Dungey53}, given by

    \begin{equation}
    B_x = \frac{B_0}{L_0} y,
    \;\;\;    B_y = \frac{B_0}{L_0} \alpha^2 x, \;\;\; 
    \label{eqn:Bxy}
    \end{equation}
where $B_0$ is characteristic magnetic field strength, $L$ is the global external length-scale of reconnection, and $\alpha$ is the stress parameter (see e.g. chapter 2.1 in Ref. \cite{priest}). A uniform current is imposed at time $t=0$ in the $z$-direction, corresponding to the curl of the magnetic field, such that Ampere's law is satisfied

  \begin{equation}
    j_z = \frac{B_0}{\mu_0 L_0} (\alpha^2 - 1).
   \end{equation}

Here, $\alpha$ greater or smaller than unity corresponds to a contraction along the $x$ or $y$ axis respectively, and results in an inwards $\vec{j}\times \vec{B}$ force on the plasma along the same axis. This in turn pushes the field lines inwards, which serves to increase the initial imbalance, which in turn increases the inwards force and the field collapses. Due to the frozen-in condition, this leads to a build up of plasma near the $X$-point and eventually to the formation of a diffusion region and a current sheet, where field lines reconnect. Since the simulation allows for kinetic effects, ions decouple, at a typical length scale of the ion inertial length ($c/\omega_{pi}$), allowing further compression of the field near the $X$-point, allowing for fast reconnection. The width of the current sheet is given by the electron inertial length, $c/\omega_{pe}$, while the length of the current sheet, which is a decisive factor in the reconnection rate, was shown to be of the order of the length of the ion diffusion region for tearing-mode reconnection \cite{shaylength}. The dominant term in the generalized Ohm's law, which results in the breaking of the frozen in condition, was shown to be the off-diagonal terms of the electron pressure tensor divergence, due to electron meandering motion (see Ref. \cite{tsiklauri2008}). This makes the reconnection process fast compared to the resistive MHD, which is too inefficient (see Ref. \cite{priest}, chapter 7.1.1).  

In all our simulation runs, $\alpha$ was set to $1.2$ and $B_0$ was adjusted such that the Alfv\'en speed at the y-boundary was fixed as $V_a=0.1c$, where $c$ is the speed of light. Further, the parameters in the simulation were adjusted to assure that the electron plasma frequency at the boundary was equal to the electron cyclotron frequency at the boundary, i.e. $\omega_{pe}=\omega_{ce}$. The number density of both electrons and ions in the simulation domain, $n_e$ and $n_p$, was set to $10^{16}m^{-2}$, while the temperature for both electrons and ions, $T_e$ and $T_p$, was set to $6.0\times10^7$K, matching conditions of flaring in the solar corona. The code was set to use 200 particles per species per cell, which was shown to be sufficient for accurate electromagnetic field dynamics in convergence tests.  
 
The simulation code used here is a 2.5D relativistic and fully electromagnetic Particle In Cell (PIC) code, developed by the EPOCH collaboration, based on the original PSC code by Hartmut Ruhl \cite{ruhl2006}. This code was modified to allow for closed boundary conditions that conserve electromagnetic flux at the boundary. To achieve this, zero-gradient boundary conditions are imposed both on the electric and magnetic fields in $x$- and $y$-directions and the tangential component of electric field was forced to zero, while the normal component of the magnetic field was kept constant. This anchors magnetic field lines on the boundary, thus inhibiting loss (or gain) of magnetic flux from the simulation domain. This is applicable to magnetic fields which are anchored in the solar photosphere by the frozen-in condition and form an X-point higher up in the corona, which serves as a simplest model of a solar coronal active region. All numerical runs in this study use closed boundary conditions (simulation results for $X$-point with open boundary conditions can be found in Refs. \cite{gvdp1,gvdp}.

\section{Octupolar structure for different domain sizes}
\pagestyle{plain}
\thispagestyle{plain}

While having the advantage of simulating a self-contained, energy conserving system, flux conserving boundary conditions naturally have a limiting effect on the movement of field lines, especially closer to the edge of the domain. In order to diminish this effect and allow natural reconnection dynamics to occur, different system sizes were used for the square simulation domain, ranging from $~4c/\omega_{pi}$ to $~32c/\omega_{pi}$. The effective grid sizes used ranged from $2L=2.14$ m to $2L=17.12$ m. For most of the runs, simulation grid cells were set to the Debye length ($\lambda_D$), thus leading to grids ranging from 400x400 to 1600x1600 grid cells. However, for the case of a system size of $~32c/\omega_{pi}$ this was not computationally possible. An under-resolved simulation run, using 1600x1600 grid cells and thus cell size of $2\lambda_D$, however proved consistent and was shown to be energy conserving despite possible numerical heating and is thus included here. The ion mass was set to 100 times the electron mass, i.e. $m_i = 100m_e$, to speed up the code. The value of $B_0$ for the different runs was adjusted such that the Alfv\'en speed at the $y$-boundary was fixed as $v_a = B_b/\sqrt{\mu_0m_in_i} = 0.1c$, where $B_b$ represents the strength of the magnetic field at $(x_{max},0)$. For meaningful comparison between the runs, the $x$-axes of plots showing time dynamics use Alfv\'en time, 
  \begin{equation}
t_a=L/v_a,
    \label{eqn:tnorm}
  \end{equation}
where $L$ denotes the half-length of the system size.

  \begin{figure}[htbp]
    \includegraphics[width=0.90\linewidth]{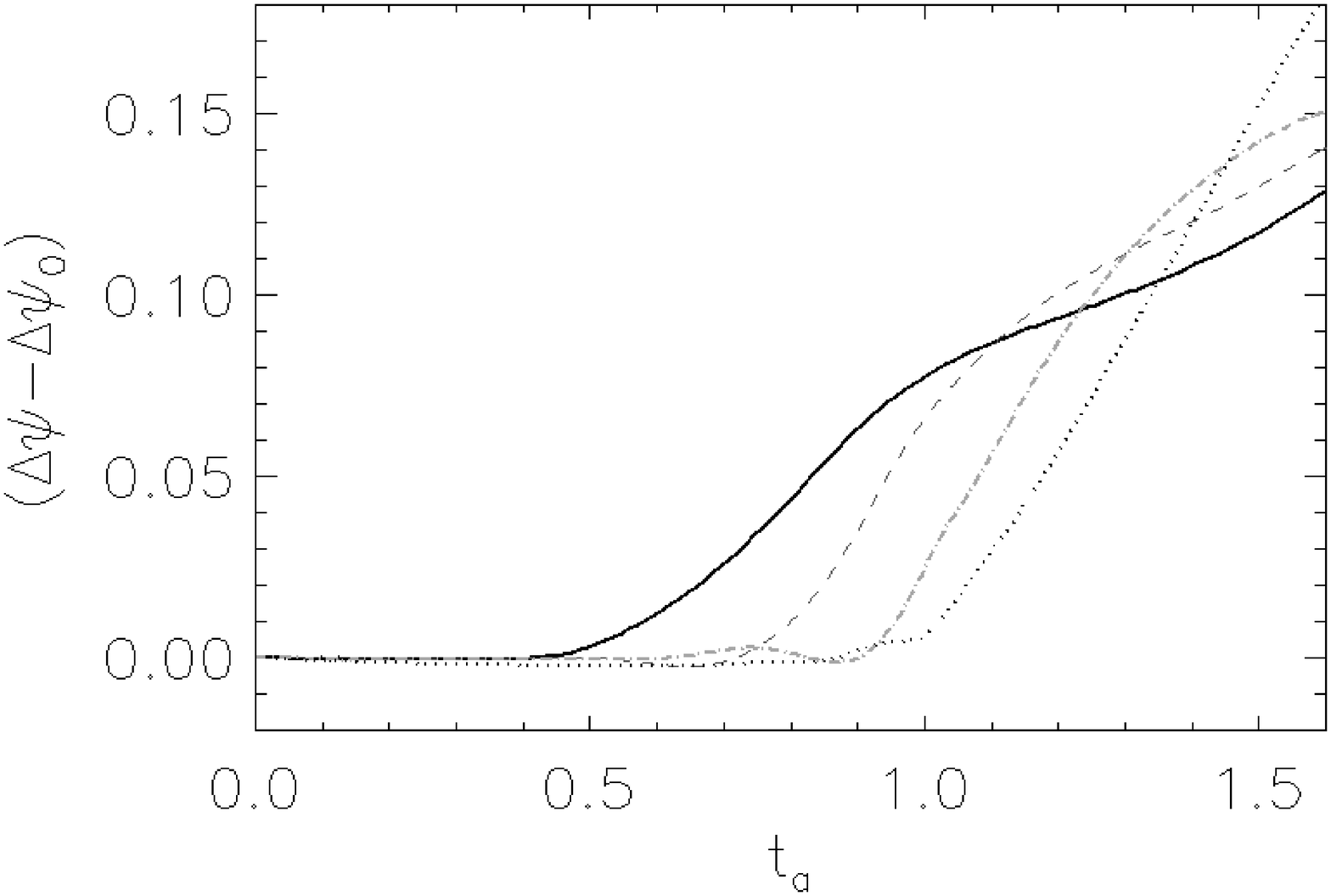}
     \caption{The reconnected flux for simulation runs of different domain sizes, given by the difference between magnetic flux at $X$-points and $O$-points, indicative of the reconnection rate (shown for different electron to ion mass ratios in Fig. 2 of Ref. \cite{tsiklauri2008}). The solid, dashed, dash-dotted and dotted curves show the reconnection rate for domain sizes of $4c/\omega_{pi}$, $8c/\omega_{pi}$, $16c/\omega_{pi}$ and $32c/\omega_{pi}$ respectively. The reconnected flux is normalised by $B_bc/\omega_{pi}$. For all system sizes a similar amount of reconnected flux is reached within $1.5t_a$. Note that, for each case, initial reconnection rate maxima are reached also within this period.}  
     \label{Fig:ezscale}
   \end{figure}

\begin{figure*}[htbp]
	\parbox[c]{.60\linewidth}{
		\includegraphics[width=\linewidth]{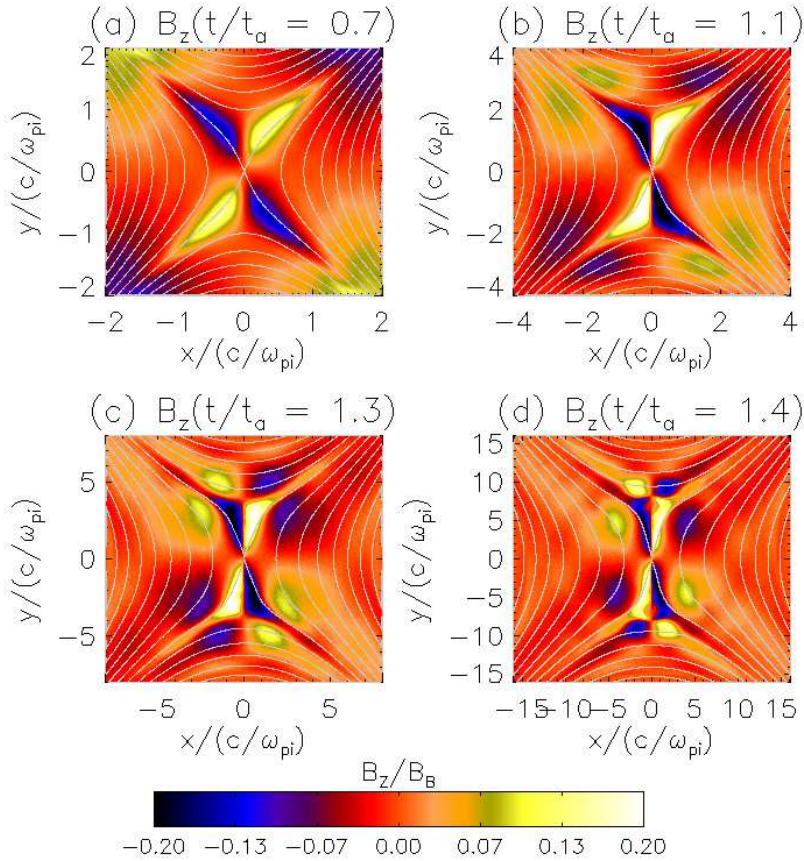}}\hfill
	\parbox[c]{.35\linewidth}{
		\caption{
The magnetic out-of-plane field for simulation runs with different domain sizes at peak reconnection rate, as indicated in Fig.~\ref{Fig:ezscale}. Panels (a) to (d) show runs using system sizes of $4c/\omega_{pi}$, $8c/\omega_{pi}$, $16c/\omega_{pi}$ and $32c/\omega_{pi}$ respectively. White lines superimposed on plots show the in-plane magnetic field at the same time. Different times correspond to time instants when maximum reconnection rate is reached. Note that $B_B$ is the same as $B_b$.}
\label{Fig:octosnaps}
	}
\end{figure*}

By plotting the reconnected in-plane magnetic flux over time for runs with different system sizes (see Fig.~\ref{Fig:ezscale}), it was determined that in all runs, a similar amount of flux is reconnected within $1.5 t_a$. For all cases an instant of maximum reconnection rate is reached within $1.5 t_a$, which occurs at progressively later times for greater system sizes (note that time normalisation is as stated in Eq. (\ref{eqn:tnorm})). This can be understood as a result of the increasing difference in length between the system size and the diffusion region. I.e. while the system size increases, the width of the diffusion region remains the same and field lines have to travel a greater distance before undergoing reconnection. For a system size of $2L=4c/\omega_{pi}$ the the peak reconnection rate is lower than for the other cases, for which the peak reconnection rate is approximately equal. In this case, the closeness of the boundary to the diffusion region is limiting the reconnection. 

The out-of-plane magnetic field for different system sizes at peak reconnection rates is plotted in Fig.~\ref{Fig:octosnaps}. The width of the out-of-plane magnetic structure does not change with the size of the domain. For system sizes of $8c/\omega_{pi}$, and greater, it can be seen that the horizontal extent of the out-of-plane field is approximately contained within $-4c/\omega_{pi}$ to $+4c/\omega_{pi}$. Since the particle density and mass ratios are fixed for the different runs, the width of ion diffusion region, $\sim c/\omega_{pi}$, is also fixed. Since ions and electrons only move independently within the ion diffusion region, which allows for in-plane currents and thus out-of-plane magnetic fields to be generated, it is evident that the emergence of octupolar structure is an aspect of ion diffusion region physics. 

The vertical extent of the out-of-plane magnetic field however does increase for greater system sizes. This shows that the generation of in-plane currents occurs at a vertical distance determined by the strength of the in-plane magnetic field, which has equal strengths on relative positions on the domain. The in-plane field lines in Fig.~\ref{Fig:octosnaps} show that the value of $B_x$ in the outflow region is fixed at approximately halfway between the $X$-point and the system boundary, for all simulation runs. Since flux can not escape from the boundary, this shows that the movement of field lines is halted at the same relative position on the domain for all system sizes. Thus, the motion of electrons must also be halted at this point since they recouple to the field shortly after reconnection. However, ions are still decoupled at this point and over-shoot, thus creating in-plane currents and charge separation. In response, electrons move along the field-lines to restore charge neutrality, resulting in the current loops and out-of-plane magnetic field as illustrated in Fig.~\ref{Fig:model} (right). In panel (a) of Fig.~\ref{Fig:octosnaps} we see that for a case of a simulation size of $2L=4c/\omega_{pi}$, the octupolar components are not contained within the domain. It is thus clear that a simulation size of at least $8c/\omega_{pi}$ is required to comprehensively model the dynamics within the ion diffusion region, without spurious influence from the boundary condition.

\section{The octupolar field structure for different ion masses}
\pagestyle{plain}
\thispagestyle{plain}

In order to determine that the results relating to octupolar structure hold true for realistic electron-to-ion mass ratios, the relationship between the strength of the octupolar and quadrupolar components and the mass ratio was investigated. In these simulation runs, the mass of electrons was fixed as their physical mass, $m_e$, and ion masses were varied from $m_i=50m_e$ to $m_i=400m_e$. In order to simulate comparable effects, runs with different ion masses were adjusted in scale, such that the domain size remained the same in number of ion inertial lengths, $c/\omega_{pi}=c\sqrt{m_i\epsilon_0/n_ie^2}$. Thus, for greater ion masses, the domain size was increased accordingly (see other findings relating to these setups in Ref. \cite{tsiklauri2008}). Further, as before, the Alfv\'en speed at the boundary was fixed ($v_a = B_b/\sqrt{\mu_0m_in_i} = 0.1c$), which means that $B_b$ was adjusted accordingly, while $B_0$ remained fixed. For meaningful comparison between the runs, time normalisations as in Eq. (\ref{eqn:tnorm}) are applied.

  \begin{figure}[htbp]
    \includegraphics[width=0.90\linewidth]{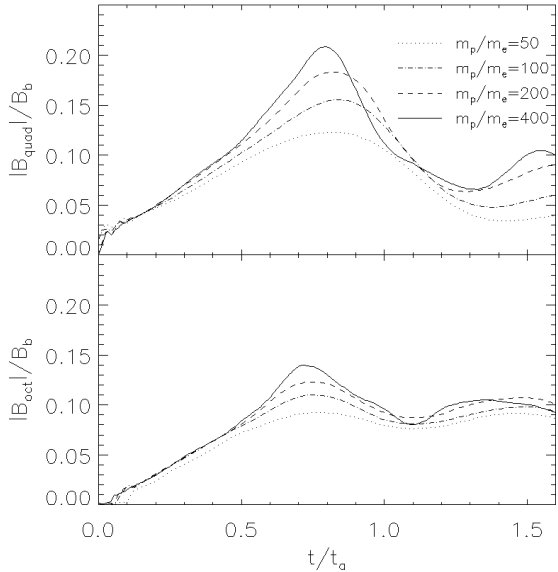}
     \caption{(Top) The strength of the quadrupolar field components for runs of different ion masses, and corresponding setups as described in the text. Different line styles correspond to different mass ratios as indicated. (Bottom) as above, showing the strength of octupolar field components.}  
     \label{Fig:bzmps}
   \end{figure}

By running the simulations described above, for a system size of $4c/\omega_{pi}$, plots in Fig.~\ref{Fig:bzmps} were obtained. Here, the strength of the qudarupolar components was determined as the maximum field strength of the quadrupolar field at each time step, and similarly for the octupolar components, i.e. max$(|B_{z,quad}(x,y)|)$ and max$(|B_{z,oct}(x,y)|)$ respectively. We gather from Fig.~\ref{Fig:bzmps} that increasing the ion mass results in a minor increase in the peak strength of the quadrupolar and octupolar field components. In all cases, the octupolar components constitute a significant fraction (i.e. $10-15\%$) of the in-plane magnetic field at the boundary and should therefore be of significance for observations in laboratory plasma experiments and spacecraft observations.

\section{Octupolar structure for tearing-mode vs $X$-point collapse}
\pagestyle{plain}
\thispagestyle{plain}

\begin{figure*}[htbp]
	\parbox[c]{.65\linewidth}{
		\includegraphics[width=\linewidth]{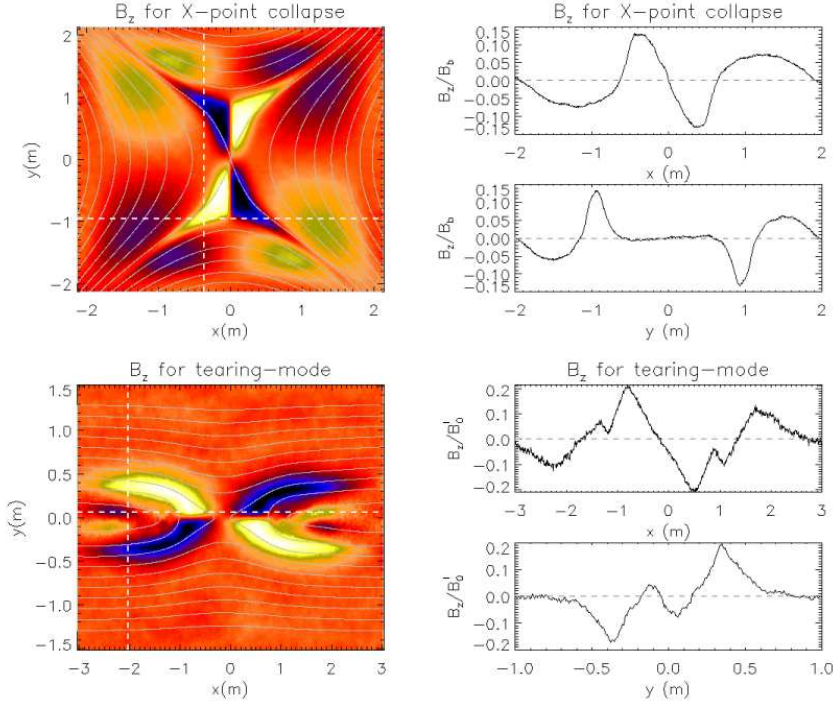}}\hfill
	\parbox[c]{.35\linewidth}{
		\caption{
(Left panels) The out-of-plane magnetic field structure at peak reconnection for $X$-point collapse ($t/t_a=1.1$) and tearing mode ($t\omega_{ci}'=20$), as in Ref. \cite{prittchet}. In both cases, elements of opposite polarity can be seen next to the inner quadrupolar structure, making an overall octupolar field. (Right panels) The line-profile of the out-of-plane magnetic field at the dotted tracks shown in the corresponding contour plots. These profiles effectively represent possible observations by a spacecraft mission, passing though reconnection regions of $X$-point collapse or tearing-mode type.}
\label{Fig:comp1}
	}
\end{figure*}

The primary factor in the emergence of octupolar structure in $X$-point collapse is the role of ion dynamics. Due to the large mass of ions compared to electrons, their contribution to the out-of-plane field has generally been considered negligible. However, by revisiting tearing-mode reconnection studies, exhibiting quadrupolar structure, it was possible for us to show that a similar octupolar structure is also present in these scenarios. Using the reconnection setup from kinetic simulation by Pritchett \cite{prittchet} for a tearing-mode reconnection, and carefully investigating the out-of-plane magnetic field structure, it was shown that octupolar components also emerge here (see Fig.~\ref{Fig:comp1}).

The initial reconnection magnetic field used in this simulation is a Harris neutral sheet configuration, given by    

    \begin{equation}
     B_x = B_0'\tanh(y/\omega), \;\;\; 
    \label{eqn:Bxprit}
    \end{equation}
together with ion and electron density profiles of 
    \begin{equation}
     n_{e,i} = n_0'\sech^2(y/\omega)+n_b, 
    \label{eqn:nprit}
    \end{equation}
where $B_0'$ is determined by the Alfv\'en speed, which is set as $v_a=c/20$. The half-thickness of the current sheet, $\omega$, is set to  $0.5c/\omega_{pi}$, $n_0'$ is determined by the equilibrium condition for the neutral sheet and $n_b$ is a constant background density of $0.2n_0'$. The mass of ions in the simulation is set as $m_i=25m_e$. By introducing an initial flux perturbation, a magnetic island with a transverse size comparable to $\omega$ is generated, leading to a reconnection region comparable to the size of the domain.

\begin{figure}[htbp]
 \includegraphics[width=0.99\linewidth]{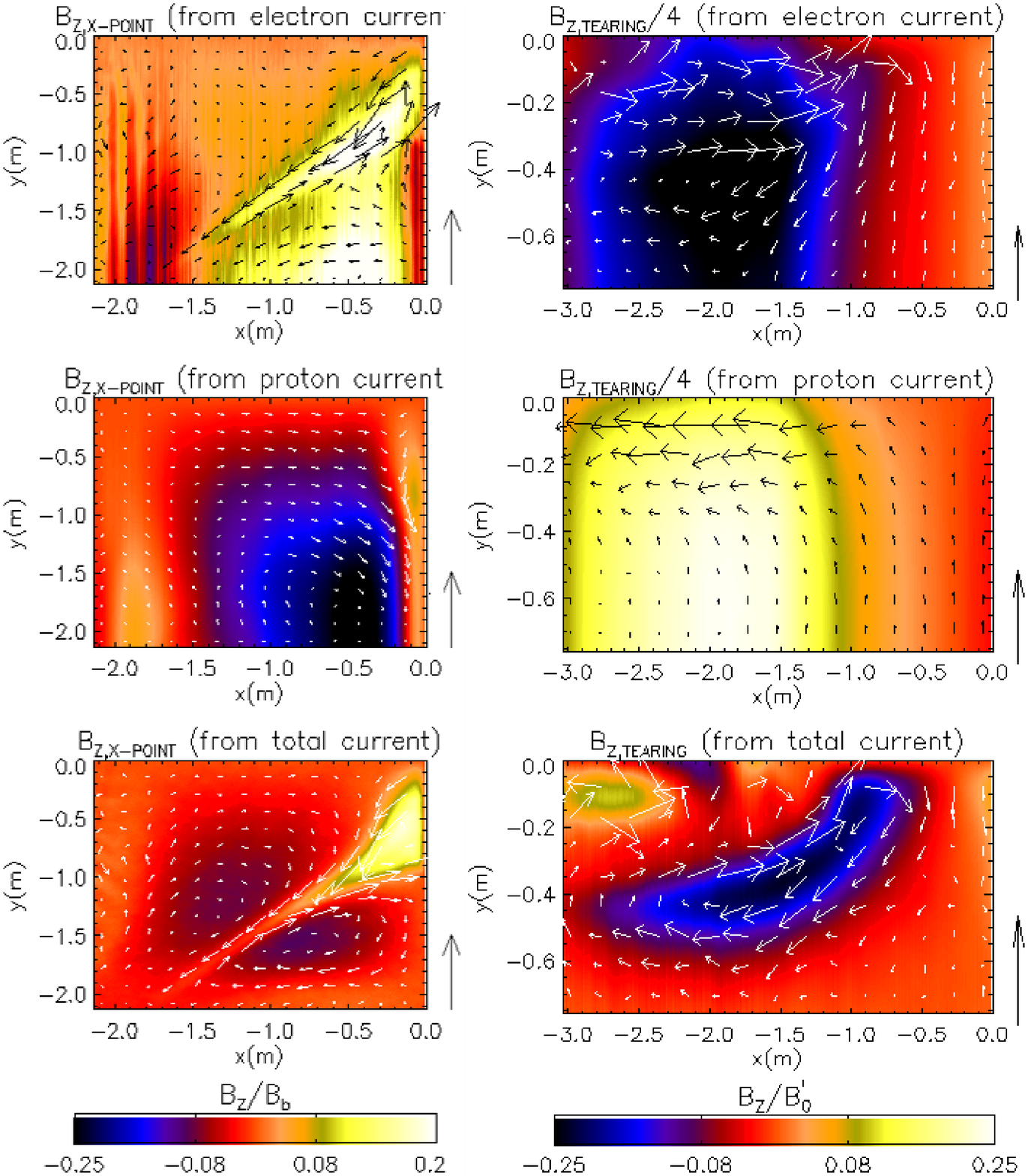}
 \caption{(Left three vertical panels) The electron and ion current contributions to the out-of-plane magnetic field for an $X$-point collapse scenario with a domain size of ~$8c/\omega_{pi}$ at peak reconnection rate ($t/t_a=1.1$), calculated from electron and ion currents respectively, based on Ampere's law. Superimposed arrows indicate the strength and direction of currents in the domain. Arrows next to plots show current strengths corresponding to charged particles of the characteristic density, moving at the Alfv\'en speed. (Right three vertical panels) The same for the tearing-mode setup ($t\omega_{ci}'=20$) according to Ref. \cite{prittchet}. Since ion and electron contributions strongly cancel, i.e. were much greater than the resulting field, different scales are used as indicated.} 
 \label{Fig:bzj}
\end{figure}

As shown in Fig.~\ref{Fig:comp1}, there are distinct differences in the out-of-plane magnetic field structure for the two reconnection setups, which could lead to different observational signatures. In $X$-point collapse, the quadrupolar field emerges in the centre of the domain and two sets of octupolar field components emerge in each of the \textit{outflow and inflow} regions. However, in the tearing-mode case, a quadrupolar field forms at the centre, but octupolar components emerge in the \textit{outflow region only}. Using line plots, the out-of-plane magnetic field along tracks through the octupolar magnetic field structure of the $X$-point collapse and tearing-mode scenarios is shown. The selected tracks, in all cases, show two consecutive troughs and peaks in the out-of-plane field. In $X$-point collapse, for both the horizontal and vertical track, the initial through and the final peak have a smaller field strength than the intermediate ones. This is the result of the track starting and ending in the outer region of the domain, where octupolar components, which are lower in field strength, dominate. In the inner region, the out-of-plane magnetic field is dominated by the quadrupolar components and greater field strengths are observed. For the tearing-mode case, for the horizontal track, this trend is also observed, but for the vertical track it is reversed, i.e. the initial trough and the final peak have a greater field strength than the intermediate ones. This can also be shown to be the result of the order in which the track passes quadrupolar and octupolar field components. In the latter case, the track passes through the quadrupolar components before and after the octupolar components, thus leading to lower field strength in the middle of the track. If a spacecraft mission, such as that discussed in Ref. \cite{Eastwood}, were to pass through a reconnection region and observe one of these line profiles, it would be possible to distinguish between the possible reconnection setups accordingly.

As in Ref. \cite{gvdp}, to determine the causes of the observed magnetic poles in terms of the currents in the simulation, Ampere's law was taken in component form, such that
    \begin{equation} 
    dB_z=\mu_0j_{x,ion}dy+\mu_0j_{x,electron}dy+\frac{1}{c^2}\frac{\partial E_x}{\partial t}dy 
        \label{eqn:ampere1}
    \end{equation}
    and
        \begin{equation} 
    dB_z=-\mu_0j_{y,ion}dx-\mu_0j_{y,electron}dx-\frac{1}{c^2}\frac{\partial E_y}{\partial t}dx.
    \label{eqn:ampere2}
    \end{equation}
    
By integrating over the domain for these terms, contributions to $B_z$ from individual currents, i.e. electron, ion and displacement current, can be calculated. The individual currents for electrons and ions are provided by the simulation at each grid cell and the displacement current was obtained by taking a five-point stencil using electric field values at the same cell, separated over four time steps. As a starting point for the integration a neutral point in $B_z$ had to be chosen. For $X$-point collapse, the most logical point was the centre of the domain, since here $B_z(0,0)=0.0$. For the tearing-mode case, any point point far out in the inflow region was suitable, as these regions are shown to be far and disconnected from the diffusion region (i.e. field lines from outside the diffusion region here do not enter the diffusion region) and thus lacking out-of-plane magnetic structure. Thus, it was possible to individually integrate over the simulation grid, using the three different currents, to obtain their individual contributions to $B_z$. E.g. using Eq. (\ref{eqn:ampere1}) one obtains

    \begin{equation} 
B_{z,ion}(0,L_y)=\int_0^{L_y} \! j_{x,ion}(0,y) \, \mathrm{d}y 
    \label{eqn:ampere3}
    \end{equation}
followed by (\ref{eqn:ampere2}) to get
    \begin{equation} 
B_{z,ion}(L_x,L_y)=B_{z,ion}(0,L_y)-\int_0^{L_x} \! j_{y,ion}(x,L_y) \, \mathrm{d}x, 
    \label{eqn:ampere4}
    \end{equation}
where $(L_x,L_y)$ represents an arbitrary point on the $B_z$ grid. 

Carrying out this integration for all $L_x$ and $L_y$ on the grid for all current contributions for both $X$-point collapse and tearing mode, plots shown in Fig.~\ref{Fig:bzj} are obtained. For considerations of symmetry, only the lower left quarter of the simulation domain is shown (also note that, since $X$-point collapse, by convention, has inflow regions along the horizontal axis and tearing-mode along the vertical axis, the sign of the quadrupolar and octupolar field structure is reversed). Top and middle panels show ion and electron current contributions to the out-of-plane magnetic field and the bottom panel the combined contributions, which lead to an out-of plane field as shown in Fig.~\ref{Fig:comp1}. As expected, based on Hall dynamics, the contribution to the quadrupolar components is provided entirely by the electron currents in both reconnection scenarios. Also, as established in Ref. \cite{gvdp}, the octupolar components are the result of ion currents. However, unlike in $X$-point collapse, in the tearing mode scenario, octupolar components only emerge in the out-flow region. Further, in the tearing mode case, due to a lack of asymmetry in the inflow region, there are two great contributions from both the ion and electron currents, which cancel to give a neutral field at the edge of the domain.

  \begin{figure}[htbp]
    \includegraphics[width=0.8\linewidth]{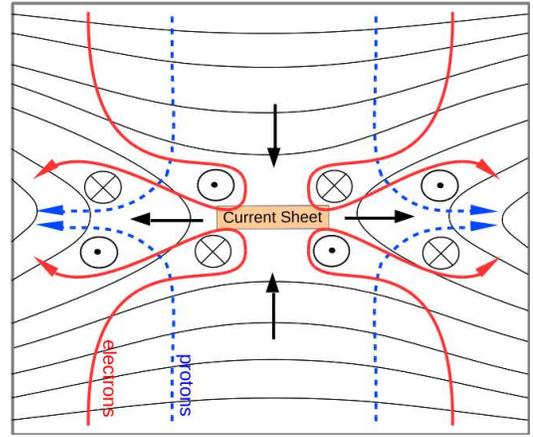}
     \caption{ \label{Fig:model2}The setup and observed resulting current and out-of-plane magnetic field generation for a tearing-mode reconnection scenario. As in Fig.~\ref{Fig:model} labelled tracks illustrate the motion of electrons and ions, generating the out-of-plane magnetic field, and black arrows on field lines indicate the inflow and outflow directions.} 
   \end{figure}

Ref. \cite{priest}, chapter 5, gives a detailed analysis of the differences in reconnection setups with almost uniform, i.e. straight inflowing field lines, and non-uniform, i.e. curved field lines, reconnection scenarios. From the results shown in this study, we find that differences in these scenarios can be extended to out-of-plane magnetic field structure. From Fig.~\ref{Fig:bzj} it can be seen that, for the tearing-mode scenario, there is a uniform current inflow of both ions and electrons in the inflow region, which is the reason no out-of-plane magnetic structure is generated in this region. This uniformity in particle inflow is a direct consequence of the lack of curvature of the inflowing field lines. In contrast to this, in $X$-point collapse, the the curvature in the field-lines means that paths of ions and electrons greatly diverge when ions decouple and thus currents and out-of-plane fields, as shown Fig.~\ref{Fig:model}, are generated. Further, from Fig.~\ref{Fig:bzj}, we can see that electrons in the tearing-mode scenario exhibit the same Hall-dynamics and flow along the field-lines towards and away from the $X$-point when entering the diffusion region. However, the field-lines in the outflow region are no longer straight and electrons move in a curved path, away from the $X$-line, while decoupled ions move in a straight horizontal path away from the $X$-point. It is this divergence in ion and electron flows which leads to the current loops that generate the octupolar components in the out-of-plane magnetic field structure in the tearing-mode case. The difference of this generation mechanism to that shown for $X$-point collapse (see Fig.~\ref{Fig:model}) can be seen in Fig.~\ref{Fig:model2}.

\section{Conclusions}
\pagestyle{plain}
\thispagestyle{plain}

Results relating to the generation of out-of-plane octupolar structure in collisionless reconnection in an $X$-point collapse \cite{gvdp} were extended by the investigation of the effect of changing domain size and ion masses. It was established that, when fixing the Alfv\'en speed at the boundary and increasing the size of the domain, the horizontal extent of the octupolar region remains approximately the same (see Fig.~\ref{Fig:octosnaps}). This was found to be consistent with previous findings since octupolar structure was shown to be the result of in-plane ion currents. As ion currents can only contribute to the out-of-plane magnetic field within the ion diffusion region and, since $c/\omega_{pi}$ remained fixed for different system sizes, the width of the octupolar region remained fixed as well. The length of the octupolar structure however increased for greater system sizes. From the in-plane field in Fig.~\ref{Fig:octosnaps} it was possible to see that the outward motion of field lines in the outflow region diminishes at the same relative position on the domain for all system sizes. As ions remain decoupled at this point, they over-shoot, causing in-plane currents. Further, electrons compensate for this by moving along the field lines, thus resulting in the current loops and out-of plane magnetic field as shown in Fig.~\ref{Fig:model}.    

By varying the electron to ion mass ratio and adjusting the simulation domain size to keep the same number of ion inertial lengths, it was shown that octupolar field strength was not significantly affected by the electron to ion mass ratio variation and consistently remained a significant fraction of the in-plane field, i.e. between $10\%$ and $15\%$ (see Fig.~\ref{Fig:bzmps}). Hence, octupolar structure in an $X$-point collapse would also be generated in plasmas with realistic mass-ratios.

Further, the discovery of octupolar structure in a tearing-mode scenario is presented and the differences in the generation process are analysed. It is found that, due to the uniform nature (i. e. straightness of field lines) of tearing-mode collapse, significant octupolar structure is only generated in the outflow region (see Fig.~\ref{Fig:bzj}). Here, ions flow out along the centre of the diffusion region, while electrons follow a curved path along the field lines (as shown in Fig.~\ref{Fig:model2}) and thus in-plane currents and out-of-plane fields are formed. Further, by analysing the line profiles of the out-of-plane magnetic field along potential tracks through the reconnection region, distinctly different profiles were obtained which could have relevance to the identification of magnetic structures found by spacecraft missions (see Fig.~\ref{Fig:comp1}).      

It is shown in Ref. \cite{hoshino}, Fig. 7, that plasma flow along the current sheet can induce streaming sausage and kink modes that generate out-of-plane magnetic fields resembling the one discussed here. Further, in Ref. \cite{dai} a framework is presented where the Hall fields of reconnection are formulated as a process of Alfv\'en eigenmode generation and dissipation. It was shown that for the $n=1$ mode that this frame work provides analytical predictions for the Hall fields, which are in good agreement with observed magnetic fields. In higher wave modes (n\textgreater1) this framework returns a higher order structure in the Hall fields which could potentially be the cause of octupolar structure in the out-of-plane magnetic field observed in this study. The application of this framework in the $X$-point collapse case is currently under investigation and will be reported elsewhere.

\acknowledgements
  Authors acknowledge use of Particle-In-Cell code EPOCH and support by development team (http://ccpforge.cse.rl.ac.uk/gf/project/epoch/). Computational facilities used are that of Astronomy Unit, Queen Mary University of London and STFC-funded UKMHD consortium at St. Andrews and Warwick Universities. JGVDP acknowledges support from STFC PhD studentship. DT is financially supported by STFC consolidated Grant ST/J001546/1 and The Leverhulme Trust Research Project Grant RPG-311.

\bibliography{ms}

\end{document}